\documentclass[twocolumn,showpacs,superscriptaddress]{revtex4-2}   
\usepackage{makecell}
\usepackage{multirow}
\usepackage{CJK}
\usepackage{graphicx}
\usepackage{mathrsfs}
\usepackage{bm}
\usepackage{amsmath}
\usepackage{dcolumn}
\usepackage{epstopdf}
\usepackage{dsfont}
\usepackage{amssymb}
\usepackage{tabularx}
\usepackage{booktabs}
\usepackage{array}
\usepackage{float}
\usepackage{color}
\usepackage{epstopdf}
\usepackage{mathrsfs}

\usepackage[colorlinks, linkcolor=blue,anchorcolor=blue,citecolor=blue,urlcolor=blue]{hyperref}
\usepackage{extarrows}
\usepackage{nicematrix}
\usepackage{tikz}
\usepackage{lineno}

\begin{document}

\title{Multiple Majorana zero modes realization based on superconducting topological crystalline metal ZrRuAs}

\author{Xiaoxu Wang}
\affiliation{Wuhan National High Magnetic Field Center $\&$ School of Physics, Huazhong University of Science and Technology, Wuhan 430074, China}
\author{Jinyu Zou}%
 \email{jyzou@hust.edu.cn}
\affiliation{Wuhan National High Magnetic Field Center $\&$ School of Physics, Huazhong University of Science and Technology, Wuhan 430074, China}%
\author{Gang Xu}%
\email{gangxu@hust.edu.cn}
\affiliation{Wuhan National High Magnetic Field Center $\&$ School of Physics, Huazhong University of Science and Technology, Wuhan 430074, China}
\affiliation{Hubei Fundamental Research Center for Physics, Wuhan, 430074, China}
\affiliation{Institute for Quantum Science and Engineering, Huazhong University of Science and Technology, Wuhan, 430074, China}
\affiliation{Wuhan Institute of Quantum Technology, Wuhan, 430074, China}

%



\date{\today}

\begin{abstract}

\section*{Abstract}
The symmetry-protected multiple Majorana zero modes (MZMs) can be manipulated under external fields and have emerged as a promising pathway toward realizing topological quantum computing. While the suitable materials hosting multiple MZMs are still scarce, we propose a feasible candidate platform named superconducting topological crystalline metals (STCMs) that simultaneously possess symmetry-protected topological bands and intrinsic superconductivity. Model analyses demonstrate that the interplay among s-wave superconductivity, mirror symmetry-protected multiple surface Dirac cones, and the introduced spin splitting leads to high BdG Chern numbers of $\mathcal{N} = \pm C_M$, where $C_M$ is mirror Chern number of the STCM. First-principles calculations identify the experimentally synthesized superconductor ZrRuAs as a promising candidate with $C_M=2$, hosting two symmetry-protected surface Dirac cones. When integrated into a heterostructure with the ferromagnetic insulator (FMI) such as GdI$_{2}$, a topological superconducting phase with $\mathcal{N} = -2$ can be realized, giving rise to two branches of MZMs. This new scheme offers advantages of structural simplicity and tunability, making the FMI/STCM heterostructure an ideal platform for investigating multipole MZMs and novel topological qubit.
\end{abstract}

\maketitle


\section{\label{sec:level1}Introduction}
Majorana fermions are fundamental particles that are equivalent to their own antiparticles~\cite{MAJORANA1937}. Although it remains unclear whether Majorana fermions exist in particle physics, they emerge in condensed matter physics as quasiparticle excitations, namely the Majorana zero modes (MZMs) in topological superconductors~\cite{Kitaev2001}. A key distinguishing feature of MZMs is their non-Abelian braiding statistics~\cite{BiaoLian2018}, which enables four MZMs to form a topological qubit. Due to the nonlocal nature of such qubits and the topological stability of MZMs, topological superconductors have become the most promising platform to realize the fault-tolerant quantum computation~\cite{DasSarma2008,10.21468/SciPostPhysLectNotes.15}. This has made topological superconductivity a highly active and prosperous field of research for over twenty years~\cite{sato2017review,Kheirkhah2020,yanPRB184505,chattervfPRB}.

However, natural topological superconductors are rare, necessitating the development of artificial strategies, among which two experimentally feasible approaches stand out. The first relies on superconductor heterostructures—including superconductor/topological insulator ~\cite{fuliang2008,BI2SE3NBSE2PRL2015} and superconductor/nanowire ~\cite{nanowirePRB024515,dasarmaPRB144522} systems—that utilize the superconducting proximity effect to induce 2D chiral or 1D topological superconductivity. However, this approach encounters difficulties such as the interfacial complexity and uncertainties in the pairing mechanism, making great challenges to the experiments. The second approach utilize the superconducting topological metals that intrinsically host both topological band structures and bulk superconductivity, such as Fe(Se,Te), in which experiments have confirmed that a magnetic vortice can host a single MZM \cite{xugang2016,Zhang2018,dinghong2018,machida2019zero,gupta2025majorana}. 

Recently, the proximity of topological crystalline insulators and superconductors are proposed to realize the multiple MZMs, which have emerged as a key focus in topological quantum computing research due to their enhanced controllability \cite{zounsr,fangchen2014,Liu2014,panPRB144501,doublePRB180505,liu2024signatures,Zhang2013,Zhang2013Kramer,Shao2026}. However, the challenge of interfacial complexity has hindered experimental realization, leaving promising material platforms scarce. This limitation motivates a critical question: can multiple MZMs be realized in superconducting topological crystalline metals (STCMs) that simultaneously possess crystalline symmetry-protected topological bands and intrinsic superconductivity? And, crucially, can we find a promising STCM candidate in nature?

In this work, we first construct a minimal model of mirror symmetry-protected STCM, and demonstrate that the interplay among s-wave superconductivity, multiple surface Dirac cones, and suitable spin splitting can realize multiple MZMs corresponding to high Bogoliubov–de Gennes (BdG) Chern number, $\mathcal{N} = \pm C_M$, where $C_M$ denotes the mirror Chern number of STCM. We then perform the first-principles calculations to identify that the experimentally synthesized superconductor ZrRuAs as an STCM with mirror Chern number $C_M = 2$, showing two Dirac cones on symmetry-preserving surface. In the GdI$_2$/ZrRuAs heterostructure, the phase diagram is determined in which topological superconducting phase with $\mathcal{N} = -2$ can be realized. Our findings establish STCMs—and particularly the ZrRuAs—as a feasible platform to explore multiple MZMs, which pave the way for the realization of topological qubit, leveraging their enhanced controllability. 

\section{Results}
\subsection{Model}
\begin{figure}
	\centering
	\includegraphics[width=0.48\textwidth]{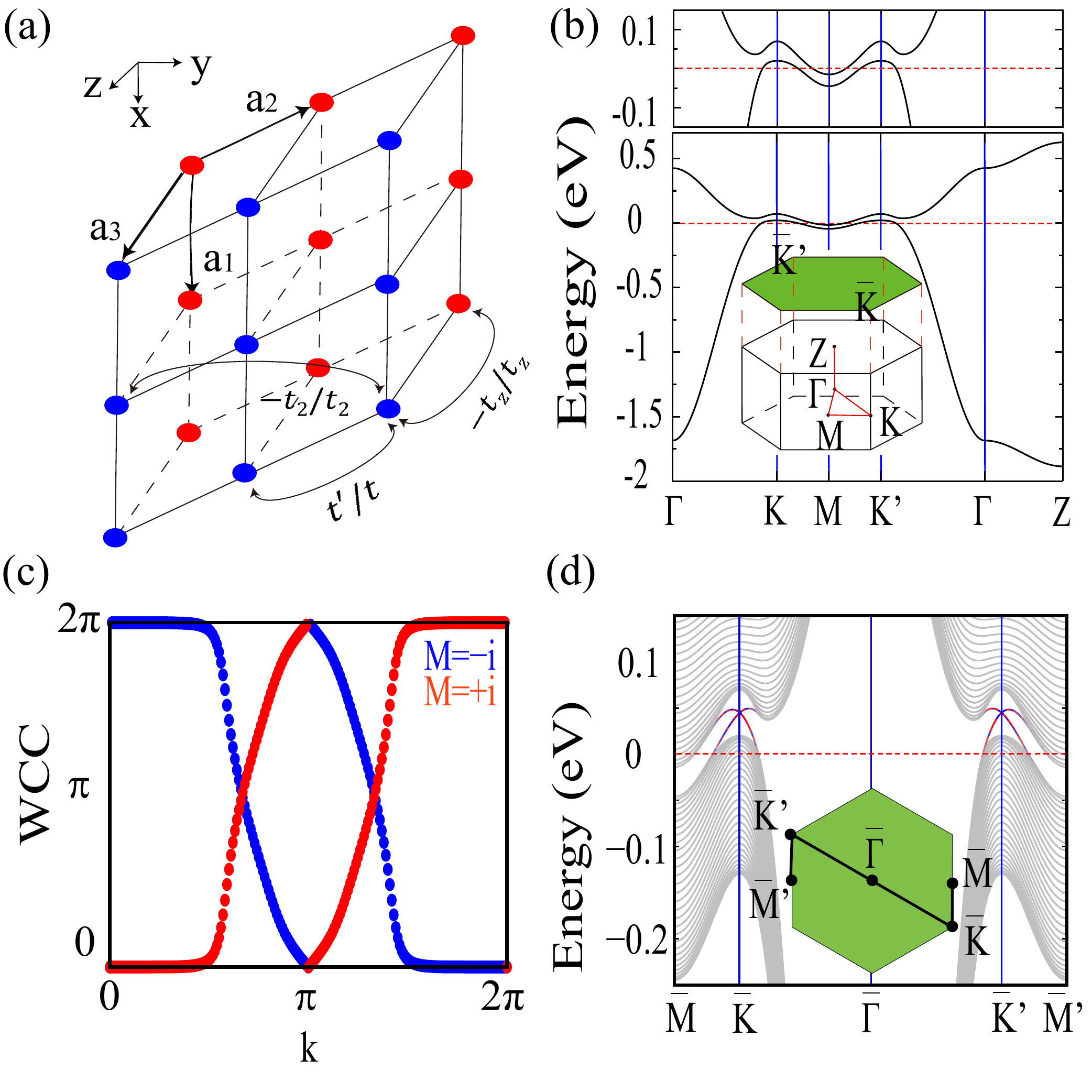}
	\caption{(a) The structure of trigonal lattice and the hopping parameters. (b) The band structure and the Brillouin zone of the bulk lattice. The parameters are set as $(m,t,t',t_2,t_z,\lambda,\lambda_z)=(-0.375,0.045,-0.195,-0.01,0.05,0.02,-0.025)$eV. The up panel is the zoom in of bands near Fermi level.  (c) The evolution of eigenvalue resolved Wannier centers~\cite{wilson2017} in the $k_{1-10}=0$ plane ($k_{1-10}=\pi$ plane is trivial and ignored here), which yields a mirror Chern number $C_{M} = 2$. (d) The 30-layer slab band structure of the topological mirror metal. The colorful bands represent the surface states.
	}
	\label{fig1}
\end{figure}

In this section,  by a minimal model, we aim to illustrate that the STCM can give rise to the chiral topological superconductivity with a high BdG Chern number, and thus realize the multiple MZMs. We first develop a 3D tight-binding model based on the trigonal lattice stacking along z-direction as illustrated in Fig.~\ref{fig1}(a), which possesses mirror symmetry $M_{1-10}$. Setting the lattice constant as unit, the lattice vectors are $a_{1}=(1 \ 0 \ 0)$, $a_{2}=(-1/2 \ \sqrt{3}/2 \  0)$ and $a_{3}=(0 \ 0 \ 1)$. Two orbits are considered on each site, denoted as $\Psi=(\psi_{1\uparrow},\psi_{1\downarrow},\psi_{2\uparrow},\psi_{2\downarrow})^T$. The in-plane nearest neighbor hopping parameter for $\psi_{1}$ ($\psi_{2}$) is $t$ ($t'$), next nearest neighbor is $-t_{2}$ ($t_{2}$), and the out-of-plane nearest neighbor  hopping parameter is $-t_{z}$ ($t_{z}$). Thus, the model is described by the Hamiltonian
\begin{align}\label{model}
H &= m \sum_{i} \Psi^\dagger_i s_{0}\sigma_{z} \Psi_i + \sum_{\langle ij \rangle} [\frac{t+t'}{2}\Psi^\dagger_i
\notag
\Psi_j + \frac{t-t'}{2} \Psi^\dagger_i s_{0}\sigma_{z} \Psi_j] \\
  &- t_2 \sum_{\langle\langle ij \rangle\rangle} \Psi^\dagger_i s_{0}\sigma_{z} \Psi_j
\notag
  +i\lambda \sum_{\langle ij \rangle} \Psi^\dagger_i (\pmb{s} \times \pmb{d}_{ij})_z \sigma_{x} \Psi_j \\
  &- t_z \sum_{\langle ij \rangle_z} \Psi^\dagger_i s_{0}\sigma_{z} \Psi_j 
  +i\lambda_z \sum_{\langle ij \rangle_z} \Psi^\dagger_i (\pmb{s} \cdot \pmb{d}_{ij}) \sigma_{x} \Psi_j
\end{align}
where $\pmb{s}$ and $\pmb{\sigma}$ are the Pauli matrices describing the spin and orbit degrees of freedom, respectively. $\langle ij \rangle$ and $\langle\langle ij \rangle\rangle$ denote the in-plane nearest and next nearest neighbor respectively, while $\langle ij \rangle_z$ denotes the out-of-plane nearest neighbor. $\pmb{d}_{ij}$ is the vector from site $i$ to $j$. $m$ is the orbit dependent onsite energy. $\lambda$ and $\lambda_z$ terms are the in-plane and out-of-plane spin-orbit coupling (SOC) respectively, which preserve the mirror symmetry $M_{1-10}$.

With adopting parameters $(m,t,t',t_2,t_z,\lambda,\lambda_z)=(-0.375,0.045,-0.195,-0.01,0.05,0.02,-0.025)$eV, a crystalline topological metal is realized as shown in Fig.~\ref{fig1}(b), which exhibits electronic pockets at $M$ and hole pockets at $K$ and $K'$. The separation between valence band and conducting band allows the topological phase protected by the $M_{1-10}$ mirror symmetry, and its mirror Chern number can be obtained by the evolution of eigenvalue resolved Wannier centers in the high-symmetry plane ~\cite{Mirrordef}. The calculations find the mirror Chern number to be $C_M=2$, as shown in Fig.~\ref{fig1}(c), indicating two mirror symmetry protected Dirac cones on the symmetry preserved surface. This is confirmed by the calculation of the energy spectra under open boundary conditions along the z-direction in Fig.~\ref{fig1}(d), which shows the surface Dirac cones at $\bar{K}$ and $\bar{K'}$. They are related by time reversal symmetry $\mathcal{T}$ and thus are described by the same low-energy 2D effective Hamiltonian  $H_D(q+\pmb{\bar{K}}/\pmb{\bar{K'}})=v_F \pmb{q}\times \pmb{s} $, where $\pmb{q}$ is the momentum near $\pmb{\bar{K}}/\pmb{\bar{K'}}$ point. Without loss of generality, we set $v_F=1$ in the following.

\begin{figure}
	\centering
	\includegraphics[width=0.48\textwidth]{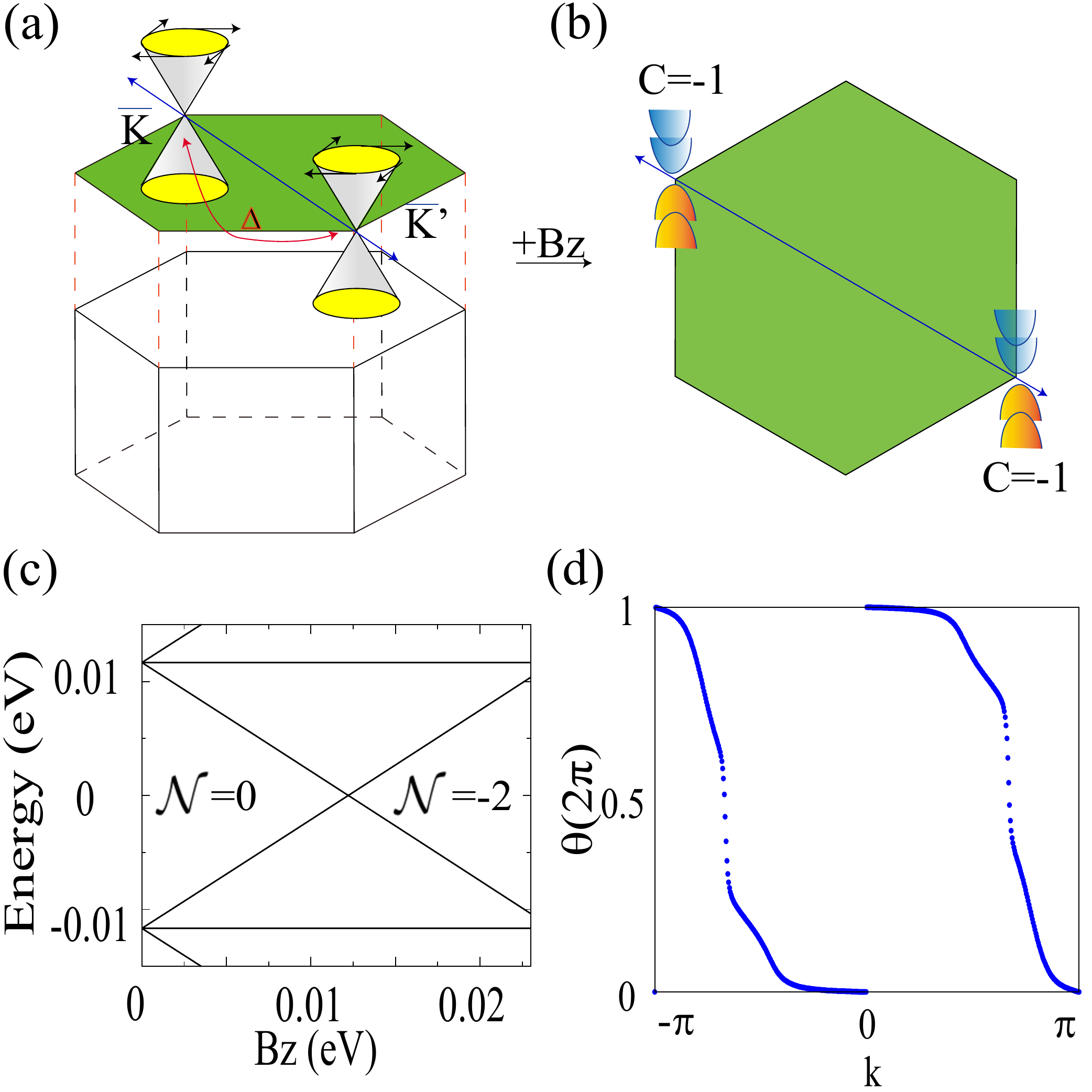}
	\caption{ (a-b) The schematic illustration that the pairing between the surface Dirac cones at $\bar{K}$ and $\bar{K'}$, associated with the spin splitting field $B_z$ can open the nontrivial gap with Chern number $\mathcal{N}=-2$. (c) The energy levels at $\bar{K}$ point as function of $B_z$, with $\Delta=10$ meV and $\tilde{\mu} = 6$ meV. (d) The Wilson loop with the same $\Delta$ and $\tilde{\mu}$, while $B_{z}$ is set as $13$ meV, yielding a BdG Chern number $\mathcal{N} = -2$.
	}
	\label{fig2}
\end{figure}

Given the metallic feature, we can consider the bulk s-wave superconducting pairing to the crystalline topological metal, making it a STCM. Due to the bulk-to-surface proximity \cite{Zhang2018,Zhang2011}, the surface Dirac cones inherit a s-wave superconducting pairing between $\bar{K}$ and $\bar{K'}$, as shown in Fig.~\ref{fig2}(a). The time-reversal symmetry $\mathcal{T}$ always eliminates the total BdG Chern number, leading to a trivial superconducting state. By introducing a $\mathcal{T}$ broken perturbation—such as an external magnetic field or ferromagnetic proximity effect induced by ferromagnetic insulator (FMI) —the surface can be driven into a topologically nontrivial phase. To illustrate such transition, we consider a spin splitting $B_z$ in addition to the s-wave pairing $\Delta$ as illustrated in Fig.~\ref{fig2}(b). Without loss of generality, we suppose $B_z>0$ and $\Delta>0$. The corresponding BdG Hamiltonian is then given by 
\begin{equation}\label{H_BdG_model}
	H^{\bar{K}\bar{K'}}_{\mathrm{BdG}}=\left( 
	\begin{array} {c|cc|c}
		H_{\pmb{\bar{K}}}(q)  & 0 & 0 & -i\Delta s_y \\ 
		\hline
		0  & H_{\pmb{\bar{K}'}}(q) & -i\Delta s_y & 0 \\
		0  & i\Delta s_y & -H_{\pmb{\bar{K}}}^*(-q) & 0 \\
		\hline
		i\Delta s_y  & 0 & 0 & -H_{\pmb{\bar{K}'}}^*(-q)
	\end{array}
	\right) ,
\end{equation}
where $H_{\pmb{\bar{K}}/\pmb{\bar{K'}}}(q)=H_D(q+\pmb{\bar{K}}/\pmb{\bar{K'}})-\tilde{\mu}s_{0}+B_{z}s_{z}$, $\tilde{\mu}$ labels the chemical potential deviating from the Dirac point. The Hamiltonian is composed of two blocks of 4$\times$4 Hamiltonian. One is the center part of Eq.~\ref{H_BdG_model} and the other is the rest. They are connected by the combined mirror $M_{1-10}$ and $\mathcal{T}$ symmetry ($M\mathcal{T}$), and thus contribute equally to the total BdG Chern number. Therefore, the topological superconducting phase of our model is protected by the combined $M\mathcal{T}$ symmetry, and we only need to study the block one of Eq. \ref{H_BdG_model} to identify the topological property. The block Hamiltonian can be expressed as
\begin{align}\label{Block}
	H_{\mathrm{block}}(q)=-q_{y}\tau_{0}s_{x}+q_{x}\tau_{z}s_{y}-\tilde{\mu}\tau_{z}s_{0}+B_{z}\tau_{z}s_{z}+\Delta\tau_{y}s_{y},
\end{align}
where $\tau$ represents the Pauli matrix for particle-hole degrees of freedom.  
We start with $\tilde{\mu}=0$ case, thus Eq.~\ref{Block} can be further block-diagonalized using a unitary transition:
\begin{align}
	U=e^{-i\frac{\pi}{4}s_{x}}e^{i\frac{\pi}{4}\tau_{x}}e^{i\frac{\pi}{4}\tau_{z}s_{x}},
\end{align}
and thus becomes 
\begin{align}\label{Blockpri}
	H_{\mathrm{block}}^{'}(q)=
	\begin{pmatrix} 
		H_{+} & 0 \\
		0 & H_{-}
	\end{pmatrix},
\end{align}
where $H_{\pm}=\pmb{q}\times \pmb{s}+(B_{z}\pm\Delta)s_{z}$ describes a massive Dirac Hamiltonian and contributes to the half Chern number $C_{\pm}=-\frac{sgn(B_{z}\pm\Delta)}{2}$~\cite{shun2018topological}. Therefore, $H_{\mathrm{block}}^{'}$, equivalently $H_{\mathrm{block}}(\tilde{\mu}=0)$ is trivial when  $B_{z}<\Delta$, while contribute BdG Chern number $\mathcal{N}=-1$ when $B_{z}>\Delta$. Given the equal contribution of the two blocks in Eq.~\ref{H_BdG_model}, the topological superconducting phase is characterized by BdG Chern number $\mathcal{N}= -2$, with the trivial phase separated by the boundary $B_{z}=\Delta$.

For the general case with $\tilde{\mu} \ne 0$, the phase boundary is determined by solving the gapless region in the energy spectra which is given by
\begin{align}
	E^2(\mathbf{k})=\Delta^{2}+q^{2}+B_{z}^{2}+\tilde{\mu}^{2}\pm2\sqrt{\Delta^{2}B_{z}^{2}+q^{2}\tilde{\mu}^{2}+B_{z}^{2}\tilde{\mu}^{2}},
\end{align}
where $q^2=q_{x}^{2}+q_{y}^{2}$. The gap closing condition $E(q)=0$ has a solution only when $|q|=0$, i.e., $B^2_{z}=\tilde{\mu}^{2}+\Delta^{2}$ which gives rise to the boundary separating the trivial and $\mathcal{N}= -2$ topological superconducting phase. 
To verify the phase boundary, we calculate the $B_{z}$ dependent spectrum at $\bar{K}$ point as show in Fig.~\ref{fig2}(c), which demonstrates an energy crossing at $B_{z} = \sqrt{\tilde{\mu}^{2}+\Delta^{2}}  \approx $12 meV, with $\tilde{\mu}$=6 meV and $\Delta$=10 meV. It indicates the BdG Chern number $\mathcal{N}=-2$ when $B_{z} > 12 $ meV, as further confirmed by the calculation in Fig.~\ref{fig2}(d) with $B_{z}=13$ meV. Our model analysis thus demonstrates that the STCM provides an ideal platform for realizing high BdG Chern number topological superconductor with multiple MZMs. 

\subsection{Candidate STCM ZrRuAs}
\begin{figure}
	\centering
	\includegraphics[width=0.48\textwidth]{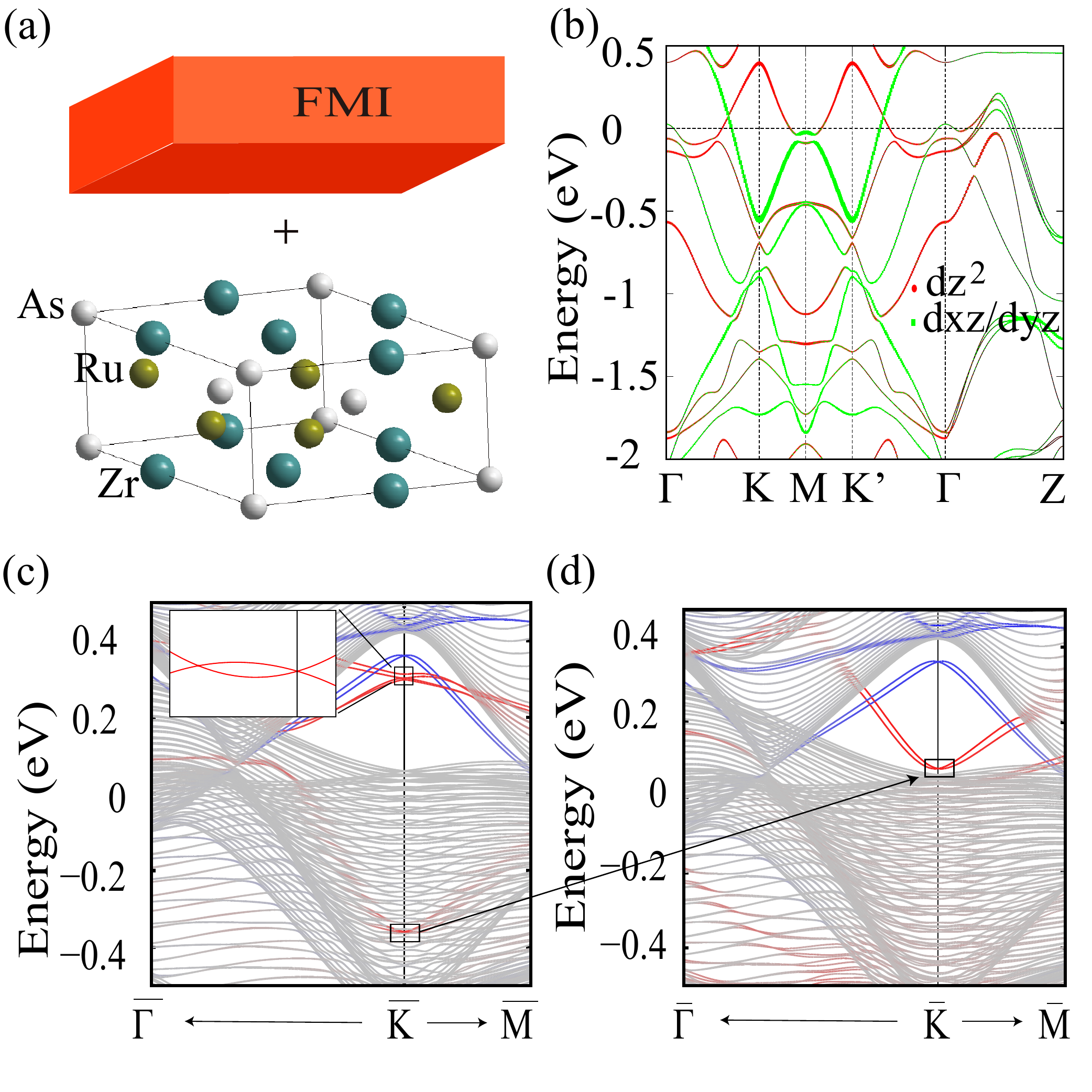}
	\caption{(a) The crystal structure of setup ZrRuAs, and the set up of heterostructure. (b) Calculation of mirror Chern number, resulting in $C_M = 2$.    (c-d) The band structure of 50-layer ZrRuAs slab. Red and blue colors represent the states localized at top and bottom surface respectively. An onsite energy of 0.55 eV is added at the top layer in (d).}
	\label{fig3}
\end{figure}

Motivated by the model analysis, we calculate and find a promising STCM candidate ZrRuAs, which is an experimentally synthesized superconductor~\cite{wangzhijun2019, MEISNER1983983}. Its s-wave superconductivity has been identified by the muon-spin rotation/relaxation ($\mu$SR) and heat capacity measurements, with a transition temperature of approximately 7.9 K~\cite{Debarchan2021} and a superconducting gap of about 1.2 meV. It crystalizes in a hexagonal primitive cell, as shown in Fig.~\ref{fig3}(a), which belongs to the space group P-62m(189) generated by rotational symmetry $C_{3_{001}}$, mirror symmetries $M_{001}$ and $M_{1-10}$\cite{wangzhijun2019}. The mirror Chern number protected by $M_{1-10}$ is $C_M=2$, as also been studied in previous research\cite{wangzhijun2019}. Fig.~\ref{fig3}(b) illustrates the band structure of ZrRuAs, which exhibits metallic feature and the band inversion at the K and K' points. Therefore, the topological bands associated with the superconductivity indicate that ZrRuAs is a promising STCM candidate.

The symmetry-protected Dirac cones on the (001) surface are shown in Fig.~\ref{fig3}(c), where the red (blue) color represents the states localized at the top (bottom) surface. $\bar{K'}$ has the same cones as $\bar{K}$, related by $\mathcal{T}$. It should be noticed that, due to the asymmetric between the top and bottom surface, their band structures are distinct. The Dirac cone of the top surface are buried in the bulk states, located around -0.35 eV, while the Dirac cone of the bottom surface are located at 0.36 eV. 

According to our model analysis, the introduction of $\mathcal{T}$ broken field can gap the surface Dirac cones and bring the topological superconducting phase with high BdG Chern number. To do that, we propose the setup of FMI/ZrRuAs heterostructure as illustrated in Fig.~\ref{fig3}(a). The FMI will bring two essential effects on the ZrRuAs slab. One is shifting the energy of surface Dirac cones due to the difference of work functions~\cite{panxiaohong2024prl,liuxinprb2023}. The other is introducing the spin splitting to the surface Dirac cones.

First, we should determine the energy position of the surface Dirac cones in the proximity to FMI. In the heterostructure of two materials with different work functions, the Fermi level of the material with higher (lower) work function will increase (decrease) to achieve equilibrium at the interface. It is equivalent to add a positive (negative) onsite energy to the surface of the material, with the value as half of the difference between the work functions \cite{panxiaohong2024prl}.
The work function of ZrRuAs on the top surface (Ru-As layer) is calculated to be 4.81 eV. The work functions of some possible FMI materials are listed in Table~\ref{tab:table1}, including LiCrSe$_{2}$, NaCrSe$_{2}$, and GdI$_{2}$. In this work, we use the GdI$_{2}$ with work function 3.71 eV. Consequently, in the GdI$_{2}$/ZrRuAs heterostructure, we can add an onsite energy of 0.55 eV on the top surface of ZrRuAs to simulate the energy shift induced by GdI2. As a result, the Dirac cones on the top surface of ZrRuAs is moved upward to 77 meV, as shown in Fig.~\ref{fig3}(d). Notably, the $M_{1-10}$ mirror symmetry is still preserved at this stage, and thereby the surface Dirac cones remain intact in the heterostructure.

\begin{figure}
	\centering
	\includegraphics[width=0.48\textwidth]{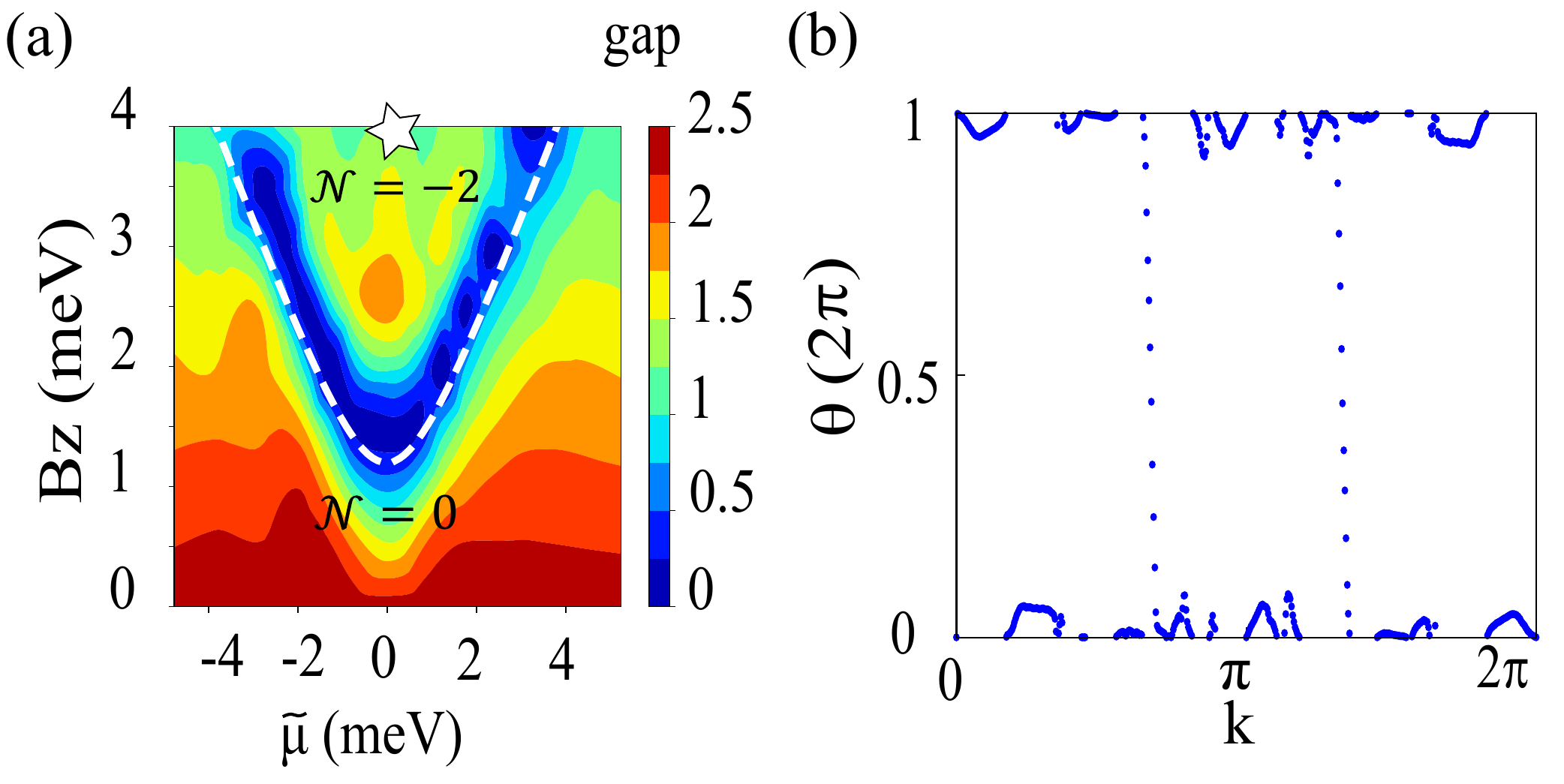}
	\caption{(a) The phase diagram of GdI$_{2}$/ZrRuAs heterostructure with $\Delta = 1.2$ meV. The color represents the minimal gap of the BdG spectrum. The white dashed line is the phase boundary determined by  $B_{z}=\sqrt{\tilde{\mu}^{2}+\Delta^{2}}$ . (b) The Wilson loop with the parameters $\tilde{\mu}=0$, $B_{z}=4$ meV, yielding a BdG Chern number $\mathcal{N} = -2$.
	}
	\label{fig4}
\end{figure}

\begin{table}
	\caption{The work function and ferromagnetic transition temperature of some candidate FMI materials.}
		\label{tab:table1}
	\begin{tabular}{|c|c|c|c|}
		\hline
		materials & LiCrSe$_{2}$~\cite{licrse22016} & NaCrSe$_{2}$~\cite{nacrse2PRM} & GdI$_{2}$~\cite{gdi21984}\\ \hline
		work function (eV) & 2.3 & 2.0 & 3.71 \\ \hline
		Tc (K) & 237~\cite{xu2020intrinsic} & 226~\cite{xu2020intrinsic} & 251~\cite{liu2021two} \\ \hline
	\end{tabular}
\end{table}
We then consider the spin splitting $B_{z}$ on the top two layers of ZrRuAs, and add the pairing gap $\Delta$ to construct the slab BdG Hamiltonian as
\begin{align}
	H_{BdG}^{slab}=
	\begin{pmatrix} 
		h^{slab}(k_{||}, B_z)-\mu & \Delta \\
		\Delta^{\dagger} &  -h^{slab}(-k_{||}, B_z)^{*}+\mu
	\end{pmatrix},
\end{align}
where $h^{slab}$ is the Wannier Hamiltonian of the ZrRuAs slab after the work function modification~\cite{hu2024numerical}, and $\mu$ is the corresponding chemical potential. For convenience in the following, we substitute $\mu$ with $\tilde{\mu}=\mu-77$ meV that labels the chemical potential deviating from the surface Dirac point.
The calculated topological phase diagram in $\tilde{\mu} - B_z$ space with $\Delta = 1.2$ meV is plotted in Fig.~\ref{fig4}(a). It is obvious that the space is divided into two regions, bounded by a nearly parabolic dark blue line indicating gap closing, which is consistent with the phase boundary determined by the theoretical formular $B_{z}=\sqrt{\tilde{\mu}^{2}+\Delta^{2}}$ (white dashed line) very well. Below the parabolic line with  $B_{z}<\sqrt{\tilde{\mu}^{2}+\Delta^{2}}$, the BdG Chern number is $\mathcal{N}=0$, while it becomes $\mathcal{N}=-2$ above the parabolic line with $B_{z}>\sqrt{\tilde{\mu}^{2}+\Delta^{2}}$, where the BdG Chern number is verified by the Wilson loop calculation as shown in Fig.~\ref{fig4}(b) with the parameters $\tilde{\mu}=0$ and $B_{z}=4$ meV (marked by the star point in Fig.~\ref{fig4}(a)). Such topological superconducting phase can be easily achieved, due to the merits that the surface Dirac points are located very closing to the Fermi level and their energy position, as well as the chemical potential, can be well controlled by the gate voltage, making ZrRuAs an excellent platform for further experimental studies.

\section{Discussion}
In summary, we propose to realize the high BdG Chern number topological superconductor based on STCMs which incorporate s-wave superconductivity, mirror symmetry-protected multiple surface Dirac cones, and the introduced $\mathcal{T}$ symmetry broken field, with the phase boundary $B^2_{z} = \tilde{\mu}^{2} + \Delta^{2}$ is derived. First-principles calculations identify the experimentally synthesized superconductor ZrRuAs as an ideal candidate material, featuring a mirror Chern number $C_M=2$. In the FMI/ZrRuAs heterostructure, the topological superconducting phase with $\mathcal{N} =-2$ can be realized by adjusting the energy position of the surface Dirac cones or chemical potential slightly. Moreover, Since the direction of the magnetization in FMI can be readily tuned, thus break the mirror symmetry of ZrRuAs, which provides additional precise control over multiple chiral MZMs evolution. This establishes the FMI/ZrRuAs heterostructure as a promising platform for investigating multiple MZMs and realizing novel topological qubits.

Finally, we would like to notice that, our study about the $C_M=2$ STCM model and material can be generalized to any $C_M$ case. As long as the  combined mirror and $\mathcal{T}$ symmetry ($M\mathcal{T}$) is preserved on the surface, the surface Dirac states can give rise to chiral topological superconducting phase with BdG Chern number $\mathcal{N} =\pm C_M$ upon the interplay between superconductivity and spin splitting. This behavior can be understood by examining the high-symmetry line of the 2D BdG Hamiltonian, which is a 1D system invariant under $M\mathcal{T}$ symmetry. Such a 1D system has been shown to belong to the BDI symmetry class and is characterized by $Z$ multiple Majorana modes at its boundaries~\cite{fangchen2014,zounsr}, which correspond to the $Z$ chiral edge modes in the parent 2D system. 

    \section{Methods}
\subsection{Methods for first-principles calculations}    
The first-principles calculations were carried out using density functional theory (DFT), which is implemented in the Vienna ab initio Simulation Package (VASP)~\cite{paw1994,pw1996,pseupot1999}. The electron-ion interaction is described by projector augmented wave (PAW)~\cite{paw1994} method. And the generalized gradient approximation (GGA) of Perdew Burke-Ernzerhof (PBE) type is adopted for the exchange-correlation
functional~\cite{pbe1996}. The internal atomic positions were fully relaxed until the forces on all atoms were reduced to below 0.01 eV/\AA. A Monkhorst–Pack k-point mesh of dimensions $5 \times 5 \times 11$ was employed. Wannier functions are obtained from Wannier90~\cite{Pizzi2020}.

\section*{Data availability}    
The data that support the findings of this study are available at https://github.com/Jinyu-Zou/ZRRUAS.\\

\section*{ACKNOWLEDGMENTS}
This work is supported by the National Natural Science Foundation of China (Grant No.~12274154, 12404182), and the National Key Research and Development Program of China (2024YFA1611200). The computation is completed in the HPC Platform of Huazhong University of Science and Technology.

\section*{Author contributions}
	
	G.X. conceived the project. X.W. performed the first-principles calculations and data analysis. X.W. and J.Z. performed the model calculation and theoretical analysis. All authors contributed to the writing of the manuscript.
	
\section*{Competing interests}
	
	The authors declare no competing interests. The authors declare no competing financial or non-financial interests.

\section*{References} 

\end{document}